\documentclass[review]{elsarticle}

\usepackage{lineno,hyperref}
\let\linenumbers\nolinenumbers\nolinenumbers

\usepackage[a4paper, total={6in, 8in}]{geometry}

\usepackage{multirow} 
\usepackage{makecell}
\usepackage{booktabs}
\usepackage{caption}
\usepackage{color}

\DeclareUnicodeCharacter{2212}{-}

\begin{document}

\begin{frontmatter}

\title{Generalisability of fetal ultrasound deep learning
models to low-resource imaging settings in five African countries}

\author[add1]{Carla Sendra-Balcells}

\author[add1]{Víctor M. Campello}
\author[add2]{Jordina Torrents-Barrena}
\author[add3]{Yahya Ali Ahmed}
\author[add4,add5]{Mustafa Elattar}
\author[add6,add7]{Benard Ohene-Botwe}
\author[add8]{Pempho Nyangulu}
\author[add8]{William Stones}
\author[add9]{Mohammed Ammar}
\author[add10]{Lamya Nawal Benamer}
\author[add11]{Harriet Nalubega Kisembo}
\author[add12]{Senai Goitom Sereke}
\author[add13]{Sikolia Z. Wanyonyi}
\author[add14]{Marleen Temmerman}
\author[add15,add16,add17]{Eduard Gratacós}
\author[add15,add18]{Elisenda Bonet}
\author[add15,add16,add17]{Elisenda Eixarch}
\author[add19]{Kamil Mikolaj}
\author[add19]{Martin Grønnebæk Tolsgaard}
\author[add1]{Karim Lekadir}

\address[add1]{Dept. de Matemàtiques i Informàtica, Universitat de Barcelona, Barcelona, Spain}
\address[add2]{HP Inc., Barcelona, Spain}
\address[add3]{Obstetrics and Gynecology Dept., School of Medicine, Suez University, Egypt}
\address[add4]{Medical Imaging and Image Processing, Center of Informatics Science, Nile University, Egypt}
\address[add5]{Research \& Development Division, Intixel, Egypt}
\address[add6]{Dept. of Radiography, School of Biomedical \& Allied Health Sciences, College of Health Sciences, University of Ghana, Ghana}
\address[add7]{Division of Midwifery and Radiography, School of Health \& Psychological Sciences, University of London, London, United Kingdom}
\address[add8]{Kamuzu University of Health Sciences, Blantyre, Malawi}
\address[add9]{Dept. of Electrical Engineering Systems \& Laboratory of Engineering System and Telecommunication, University of M'Hamed Bougara Boumerdes, Algiers, Algeria}
\address[add10]{Obstetrics and Gynecology Dept., School of Medicine, Algiers University, Algeria}
\address[add11]{Dept. of Radiology, Mulago National Referral and Teaching Hospital, Kampala, Uganda}
\address[add12]{Dept. of Radiology and Radiotherapy, School of Medicine, Makerere University College of Health Sciences, Kampala, Uganda}
\address[add13]{Dept. of Obstetrics and Gynaecology, Aga Khan University Hospital, 3rd Parklands Avenue, Nairobi, Kenya}
\address[add14]{Centre of Excellence in Women and Child Health, Aga Khan University, Nairobi, Kenya}
\address[add15]{BCNatal Fetal Medicine Research Center, Hospital Clínic and Hospital Sant Joan de Déu, Universitat de Barcelona, Barcelona, Spain}
\address[add16]{Institut d'Investigacions Biomèdiques August Pi i Sunyer (IDIBAPS), Barcelona, Spain}
\address[add17]{Centre for Biomedical Research on Rare Diseases (CIBER-ER), Barcelona, Spain}
\address[add18]{Barcelona Tech, Universitat Politècnica de Catalunya, Barcelona, Spain}
\address[add19]{Copenhagen Academy for Medical Education and Simulation and Dept. of Obstetrics, Rigshospitalet, Denmark}
\address{*Corresponding authors. Email addresses: carla.sendra@ub.edu (Carla Sendra-Balcells), victor.campello@ub.edu (Víctor M. Campello)}

\begin{abstract}
Most artificial intelligence (AI) research and innovations have concentrated in high-income countries, where imaging data, IT infrastructures and clinical expertise are plentiful. However, slower progress has been made in limited-resource environments where medical imaging is needed. For example, in Sub-Saharan Africa, the rate of perinatal mortality is very high due to limited access to antenatal screening. In these countries, AI models could be implemented to help clinicians acquire fetal ultrasound planes for the diagnosis of fetal abnormalities.
So far, deep learning models have been proposed to identify standard fetal planes, but there is no evidence of their ability to generalise in centres with low resources, \emph{i.e.} with limited access to high-end ultrasound equipment and ultrasound data. This work investigates for the first time different strategies to reduce the domain-shift effect arising from a fetal plane classification model trained on one clinical centre with high-resource settings and transferred to a new centre with low-resource settings. To that end, a classifier trained with 1,792 patients from Spain is first evaluated on a new centre in Denmark in optimal conditions with 1,008 patients and is later optimised to reach the same performance in five African centres (Egypt, Algeria, Uganda, Ghana and Malawi) with 25 patients each.
The results show that a transfer learning approach for domain adaptation can be a solution to integrate small-size African samples with existing large-scale databases in developed countries. In particular, the model can be re-aligned and optimised to boost the performance on African populations by increasing the recall to $0.92 \pm 0.04$ and at the same time maintaining a high precision across centres.
This framework shows promise for building new AI models generalisable across clinical centres with limited data acquired in challenging and heterogeneous conditions and calls for further research to develop new solutions for the usability of AI in countries with fewer resources and, consequently, in higher need of clinical support.
\end{abstract}


\end{frontmatter}

\linenumbers

\section{Introduction}

In the age of Big Data and digital health, great excitement has been generated around the extraordinary opportunities that artificial intelligence (AI) may offer in tomorrow's healthcare. In particular, deep learning-based methods have shown great promise for the analysis of complex biomedical data. In medical imaging, deep learning has already made an impact in a wide range of applications such as cardiology~\cite{martin2020image}, cancer~\cite{houssein2021deep} or brain imaging~\cite{zhang2020survey} among many others~\cite{zhou2021review,aggarwal2021diagnostic}. However, existing developments have been mostly focused on applications in high-resource settings, where there is greater access to large clinical imaging datasets for training deep learning networks. Applications in environments with limited resources, such as countries across the African continent, have been scarce. Thus, there is a risk of increasing inequalities in global health due to the disparities in the application of AI in medical imaging.

Paradoxically, in places with important medical imaging needs the development of AI is at its lowest level. For example, in Sub-Saharan Africa, there is a very high level of neonatal mortality (\emph{i.e.} 27 deaths per 1,000 births in 2019, compared to an average rate of 3.4 per 1,000 in the European Union~\cite{who_mortality}) and a very high burden of stillbirths~\cite{de2016stillbirths}, but the access to antenatal screening is limited, especially in rural Africa. A statistical study from the World Health Organisation highlights a critical shortage of trained obstetricians across African countries~\cite{who_obstetricians}. However, the diagnosis of fetal abnormalities in clinical practice requires clinical expertise to acquire and interpret ultrasound images of the fetus.

At the same time, the literature is abundant with AI developments and deep learning implementations to facilitate the tasks of fetal ultrasound scanning, quantification and diagnosis \cite{torrents2019segmentation, liu2019deep, komatsu2021detection, matthew2022exploring, plotka2021fetalnet, yin2021ultrasonographic}. For example, several techniques have been developed to help clinicians acquire fetal ultrasound planes that are adequate for further quantification, \emph{i.e.} that cover specific structures such as the fetal head, fetal femur, fetal abdomen and fetal thorax. 
Baumgartner et al. \cite{baumgartner2017sononet} were the first to identify and localise 13 fetal standard planes in real time during 2D ultrasound mid-pregnancy examinations using convolutional neural networks (CNNs). Later, similarly but in a retrospective manner, Maraci et al. \cite{maraci2017framework} automated the task of detecting the fetal presentation and heartbeat from a pre-defined free-hand ultrasound video with support vector machines and a conditional random field model instead of CNNs, due to the limited number of samples available for their study. Alternative approaches involved obtaining standard planes from 3D ultrasound (US) volumes to learn the mapping between the 2D plane and the transformation required to move towards the standard plane within the 3D volume \cite{li2018standard}. More recently, Burgos-Artizzu et al.~\cite{burgos2020evaluation} evaluated the maturity of state-of-the-art CNNs to automatically classify 2D maternal fetal US and released a large open-source dataset to promote further research on the matter. While some AI tools are being implemented to recognise multiple views in fetal imaging \cite{ge}, others focus their work on specific anatomical structures \cite{montero2021generative, wang2021recognition, sushma2021classification}.  

Despite the big amount of research studies, the proposed techniques have been trained and/or tested with fetal ultrasound datasets from high-income countries, such as the UK~\cite{baumgartner2017sononet, maraci2017framework, li2018standard}, China~\cite{wang2021recognition}, United States~\cite{ge} or Spain \cite{burgos2020evaluation, montero2021generative}. There is no evidence of the ability of these deep learning models to generalise in centres with low resources, \emph{i.e.} with lower image quality and a limited number of data to calibrate the models. To our knowledge, only one study assessed the applicability of a fetal ultrasound plane classifier across countries, but exclusively in high-resource imaging settings: a classifier was trained in the UK and tested successfully in Denmark~\cite{tolsgaard2021does}.

\begin{figure}[hbt!]
\centering
\includegraphics[width=\textwidth]{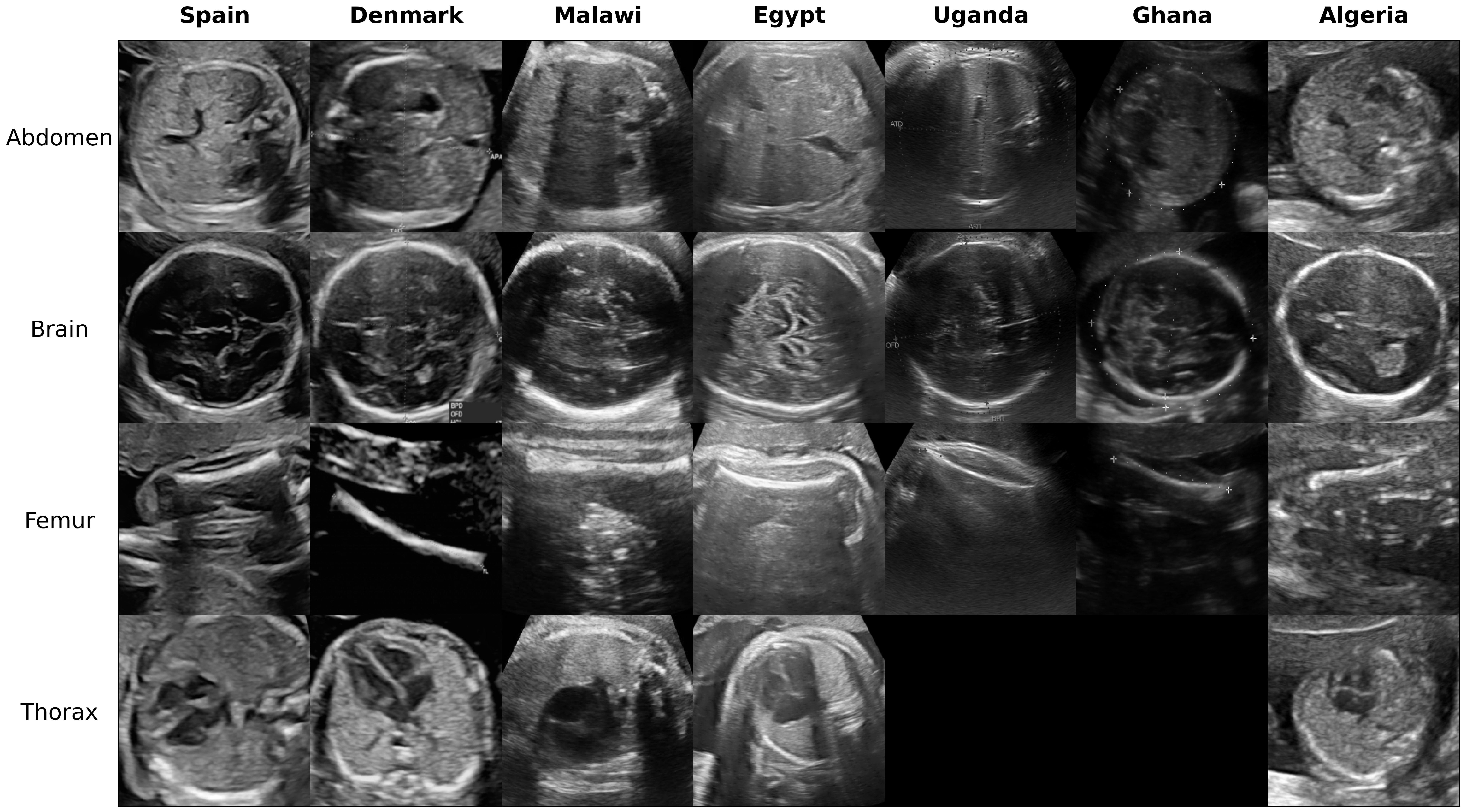}
\caption{Image examples of the maternal-fetal US categories from our multi-centre dataset.}
\label{fig:datasets}
\end{figure}

This paper presents the very first study investigating the transferability of AI models trained in high-income settings and their applicability to process images acquired in low-income settings. Specifically, we build fetal ultrasound classifiers using large-scale datasets from Spain and assess their performance when applied to fetal ultrasounds acquired in five different African countries with differences in scanners, populations and contexts. Furthermore, transfer learning, an effective strategy to adapt a pre-trained model to a new domain \cite{kim2022transfer}, is implemented and assessed under real-world conditions, \emph{i.e.} based on very small size samples of new images from the local sites, to assess the resources and efforts needed to re-calibrate the original model to obtain high-performance in each local clinical centre. The main contributions of this work are summarised as follows:

\begin{itemize}
  \item The collection and annotation of a new multi-centre fetal ultrasound dataset comprised of 120 patients from five African countries, namely Algeria, Egypt, Malawi, Uganda and Ghana. This dataset is the most diverse ultrasound dataset acquired to our knowledge and represents real clinical scenarios across low-resource imaging settings. As a result, the ultrasound appearance, both globally and locally, exhibits marked differences as clearly illustrated in Figure \ref{fig:datasets}.
  \item Application and evaluation of a fetal plane classification model trained in high-income settings within a diversity of clinical centres in low-resource imaging settings. To that end, a deep learning model is built using a large and diverse European dataset from Spain (12,400 images from two hospitals) \cite{burgos2020evaluation} and externally evaluated in two different settings: 1) with fetal ultrasounds from four centres located in another European country, namely Denmark \cite{tolsgaard2021does}, to assess generalisability under similar settings and resources; and 2) with the fetal ultrasound images acquired from five African countries to assess generalisability under different conditions and contexts.
  \item Implementation of adaptive learning approaches to optimise the scalability of the classifiers in settings with distinct imaging conditions and lower image quality. This includes the estimation of the minimal number of samples required for calibrating the models to reach maximal performance for standard fetal plane classification in each of the local African centres. As a result, realistic strategies are derived for the scalability and applicability of existing deep learning models across new centres and imaging settings with minimal effort.
\end{itemize}

\section{Methods}

\subsection{Datasets}

The data used in this study and described in Table~\ref{tab:dataset acquisition}, consist of seven datasets acquired in distinct clinical centres. The first is a publicly available dataset from two centres in Spain~\cite{burgos2020evaluation}. The second has been acquired in four different centres in Denmark. Finally, the last five datasets contain US samples from five African countries (Malawi, Algeria, Uganda, Ghana and Egypt). All datasets were fully anonymised according to institutional guidelines. The Spanish dataset is publicly available, so no approval was necessary for this work. Subjects in Denmark, Ghana and Algeria signed informed consents and the studies were conducted under the approval of the ethics committee from each scan centre. The committee responsible in Denmark was the Danish Patient Safety Authority and the Danish Data Protection Agency. For Ghana, the Ethics and Protocol Review Committee of the School of Biomedical \& Allied Sciences, University of Ghana. And for Algeria, the Ethics Committee of the Kouba Hospital, Algeria. For Malawi, Egypt and Uganda datasets, no individual consent was necessary according to national guidelines since no identifiable data was collected. The subjects have been scanned using a range of scanner US vendors by GE Medical Systems, Siemens, Edan Instruments, Shenzhen Mindray Bio-Medical Electronics and Aloka. Moreover, the frequency range of the curved transducer used during the acquisition ranges from 3 to 7.5 MHz and the screening was performed during the 2nd and/or 3rd trimester of the pregnancy period. Finally, clinical experts or research technicians who received extensive training in US classification were asked to classify all images into femur, thorax, head and abdomen. The other less common planes were classified into the `other' category. However, the five African datasets lack the `other' category, and two of them do not contain samples of the thorax, representing a real clinical scenario in low-resource settings. Details about the number of samples used in each case are summarised in Table \ref{tab:dataset numbers}. 

\renewcommand{\arraystretch}{1.5}
\begin{table}
    \captionof{table}{Details about the seven datasets considered in the study.}
    \scriptsize
    \centering
    \begin{tabular}{ |p{1.1cm}|p{4.3cm}|p{1.5cm}|p{1.cm}|p{3.cm}|p{1.45cm}|}
    \hline
    \textbf{Country} & \textbf{Vendors} & \textbf{Type of transducer} & \textbf{Freq range (MHz)} & \textbf{Name of clinical centre}& \textbf{Trimester pregnancy}\\ 
    \hline
    Spain & Voluson E6, Voluson S8 and Voluson S10 (GE Medical Systems, Zipf, Austria), and Aloka (Aloka CO., LTD.) & Curvilinear transducer & 3 to 7.5 & Hospital Clínic and Hospital Sant Joan de Déu & 2nd and 3rd\\
    \hline
    Denmark & Voluson E6, Voluson S8 and Voluson S10 (GE Medical Systems, Zipf, Austria) & Curvilinear transducer & 3 to 7.5 & Copenhagen University Hospital Rigshospitalet, Hvidovre Hospital, Herlev Hospital and Nordsjællands Hospital Hillerød. & 2nd and 3rd\\
    \hline
    Malawi & Mindray DC-N2 (Shenzhen Mindray Bio-Medical Electronics Co., Ltd, China/Germany) & Curvilinear transducer & 3.5 & Queen Elizabeth Central Hospital & 2nd and 3rd\\ 
    \hline
    Egypt & Voluson P8 (GE Medical Systems, Zipf, Austria) & Curvilinear transducer & 7 & Sayedaty Center & 2nd\\
    \hline
    Uganda & ACUSON X600 (Siemens) & Curvilinear transducer & 3 to 7.5 & Mulago National Referral Hospital & 3rd\\
    \hline
    Ghana & EDAN DUS 60 (Edan Instruments, Inc., Shenzhen,  China) & Curvilinear transducer & 3.5 to 5 & KBTH Polyclinic (Accra) & 2nd and 3rd\\
    \hline
    Algeria & Voluson S8 (GE Medical Systems, Zipf, Austria) & Curvilinear transducer & 3 to 7.5 & EPH Kouba and Clinique Des Lilas & 2nd and 3rd\\
    \hline
    \end{tabular}
\label{tab:dataset acquisition}
\end{table}

\subsubsection{European datasets}

The first large and open-source dataset with 1,792 patients was collected at BCNatal~\cite{burgos2020evaluation}, an institution with two associated hospitals (Hospital Cl\'inic and Hospital Sant Joan de D\'eu, Barcelona, Spain). All pregnant women attending routine pregnancy screening during the 2nd and 3rd trimesters between October 2018 and April 2019 were included in the study. Images were acquired using six different US machines by different operators with similar experience. The US machines used were three Voluson E6, one Voluson S8, one Voluson S10 (GE Medical Systems) and one Aloka. Images were taken using a curved transducer with a frequency range between 3 and 7.5 MHz. Operators were instructed to avoid using any type of post-processing or artefacts such as smoothing, noise, pointers or callipers when possible.

The second big dataset of this study was composed of 2nd and 3rd trimester US scans obtained from four fetal medicine centres in Denmark (Copenhagen University Hospital Rigshospitalet, Hvidovre Hospital, Herlev Hospital and Nordsjællands Hospital Hillerød). The scans were completed in the period between 2009 and 2017 and were obtained using three GE machines: Voluson E7, Voluson E8 and Voluson E10. Images were taken using a curved transducer with a frequency range between 3 and 7.5 MHz and no post-processing or artefacts were applied. 

\subsubsection{African datasets}

Five African datasets were collected specifically for this study between November 2021 and February 2022. Each dataset included 25 patients and the images were not processed after the acquisition.

The first African dataset was acquired at Queen Elizabeth Central Hospital in Malawi using a Mindary DC-N2 US machine. The 2nd and 3rd trimester US samples were obtained between December and February using a curved transducer with a 3.5 MHz frequency. 

The second small-size dataset was obtained at Sayedaty centre in Egypt using a Voluson P8 (GE), a GE Voluson series ultrasound with less image quality. The 2nd-trimester ultrasound scans were acquired between November and December using a curved transducer with a 7 MHz frequency. 

The third African dataset consists of images from pregnant women during the 3rd trimester from the Mulago National Referral Hospital in Uganda. The US images were obtained in December 2021 with a Siemens US machine and using a curved transducer with a frequency range between 3 and 7.5 MHz. 

The fourth dataset was acquired with an EDAN DUS 60 US machine and comprises samples from women in their 2nd or 3rd trimester pregnancy period. The images were acquired at the KBTH Polyclinic Centre, Accra, Ghana, in December. Operators used a curved transducer with a frequency range between 3.5 and 5 MHz. 

The last dataset was acquired between November 2021 and December 2021 at the EPH Kouba and Clinique Des Lilas centres in Algeria and using the same acquisition as in the European datasets, but with a lower-quality GE machine.

\renewcommand{\arraystretch}{1.5}
\begin{table}
    \captionof{table}{Number of subjects for each of the four datasets used during the training, validation and testing phases.}
    \scriptsize
    \centering
    \begin{tabular}{ |p{1.25cm}|p{1.5cm}|p{1.cm}|p{1.cm}|p{1.cm}|p{1.cm}|p{1.cm}|p{1.25cm}|}
    \hline
    \textbf{Country} & \textbf{Abdomen} & \textbf{Brain} & \textbf{Femur} & \textbf{Thorax}& \textbf{Other}& \textbf{TOTAL}& \textbf{Patients}\\ 
    \hline
    Spain & 711 & 1781 & 1040 & 1718 & 4213 & 9463 & 1792\\
    \hline
    Denmark & 771 & 635 & 844 & 291 & 0 & 2541 & 1008\\
    \hline
    Malawi & 25 & 25 & 25 & 25 & 0 & 100 & 25 \\ 
    \hline
    Egypt & 25 & 25 & 25 & 25 & 0 & 100 & 25\\
    \hline
    Uganda & 25 & 25 & 25 & 0 & 0 & 75 & 25\\
    \hline
    Ghana & 25 & 25 & 25 & 0 & 0 & 75 & 25\\
    \hline
    Algeria & 25 & 25 & 25 & 25 & 0 & 100 & 25\\
    \hline
    \end{tabular}
\label{tab:dataset numbers}
\end{table}

\subsubsection{Data preparation and pre-processing}

All patient data was fully anonymised by removing the header in the original image and was later stored in PNG (Portable Network Graphic) format without compression to avoid quality degradation. Since US images do not contain colour information, images were stored as grayscale bitmaps. Images with visual artefacts were flagged by the clinician or technician and excluded from the analysis. We cropped the region containing most of the field of view but excluded the vendor logo and ultrasound control indicators. Finally, a min-max normalization was used after resizing the image to 224x224 to keep the same intensity range in images from the same dataset.

\subsection{Model architecture and implementation details}

Based on the model benchmark presented by Burgos et al.~\cite{burgos2020evaluation} for fetal plane classification, the best performing method, a Densenet-169~\cite{huang2017densely}, was implemented in this study. The Spanish dataset described previously was used as the training set for the aforementioned benchmark, achieving a 0.94 average class precision. The Adam optimiser was used for training the model with a constant learning rate of $10^{-3}$ and first and second moments of 0.9 and 0.999, respectively. At each iteration, a batch of size 24 was used to calculate the loss function (cross-entropy loss) and optimise the network parameters. To improve the learning of the model, data augmentation was implemented during training. At each batch, images were randomly flipped, cropped between 0--20\%, translated from 0--10 pixels and randomly rotated up to an angle of 15 degrees. The code was developed in PyTorch \cite{paszke2019pytorch} and the hardware used during training and inference consisted of an 8 GB NVIDIA GeForce RTX 2080 Ti GPU. The different datasets were split into 50\% for training and 50\% for testing because of the small sample size of the African datasets. Different patients were included in each split to avoid data leakage. During training, 20\% of the studies were kept for validation for those datasets with a large sample size case, while this amount was increased to 50\% in the small sample size case. The same class distribution as in the Spanish dataset was enforced in the remaining datasets (40\% `other' and 15\% for each of the other planes: femur, thorax, abdomen and head) and the missing categories of the African datasets (`other' or thorax) were filled with samples from the Spanish dataset.

\subsection{Ablation study}

Several strategies were considered to study the model transferability from high to low-resource imaging settings. We classified them into single-centre and multi-centre. The first approach uses a single centre in high-resource settings and the second approach includes also samples from the target centre in low-resource settings. 

\begin{enumerate}
  \item \textbf{Single-centre}: We train the neural network with the Spanish dataset using a different amount of samples ($n=125$, 250, 500, 1000, 2000, 4000). Then, we evaluate the classification performance on the new datasets to test the generalisability of the classification model either on the Danish dataset, acquired in high-resource settings or on the African datasets acquired in low-resource settings.
  
  \item \textbf{Multi-centre}:
  \begin{itemize}
      \item[] \textbf{Combination}: We train the neural network directly using two samples with varying sizes from two different datasets, the Spanish dataset ($n=0$, 125, 250, 500, 1000, 2000, 4000) and each of the African datasets separately ($n=25$). With this, we investigate the effect of the sample size of the first centre on the performance of the model in the second centre.
      \item[] \textbf{Transfer learning}: We fine-tune the last 4 layers of a pre-trained neural network with a few US images from the new clinical site in Africa. To that end, we pre-train the neural network with the Spanish dataset using again a different amount of samples ($n=0$, 125, 250, 500, 1000, 2000, 4000), being $n=0$ the case in which a model pre-trained on the public ImageNet database for object detection and image classification is used. Then, the model is fine-tuned with the available US samples from each of the African centres. Additionally, we estimate the minimum number of patients ($p=2$, 4, 6, 8, 10, 12) needed from the African centre during the fine-tuning to obtain the desired level of performance, compared to the case in which no pre-trained model is used.
  \end{itemize}
\end{enumerate}

\subsection{Performance evaluation}

The average accuracy metric is used for monitoring the evolution of the model training. Thus, the model that reaches the maximum accuracy on the validation data and the minimum validation loss is saved as the final model and is later assessed against the testing set. For all experiments and results, confusion matrices are calculated and saved. The area under the curve (AUC) score, a metric that did not influence the training of the model, is used to assess all the experiments. Since the classification involved multi-label outputs, the AUC was computed as the average of the scores for one-versus-rest comparisons, \emph{i.e.} AUC scores were computed taking each label as the positive class and all the rest as negatives.

\section{Results}

\subsection{Generalisation of the fetal plane classifier to high or low-resource settings using a single-centre approach}

\begin{figure}[h]
\centering
\includegraphics[width=0.5\textwidth]{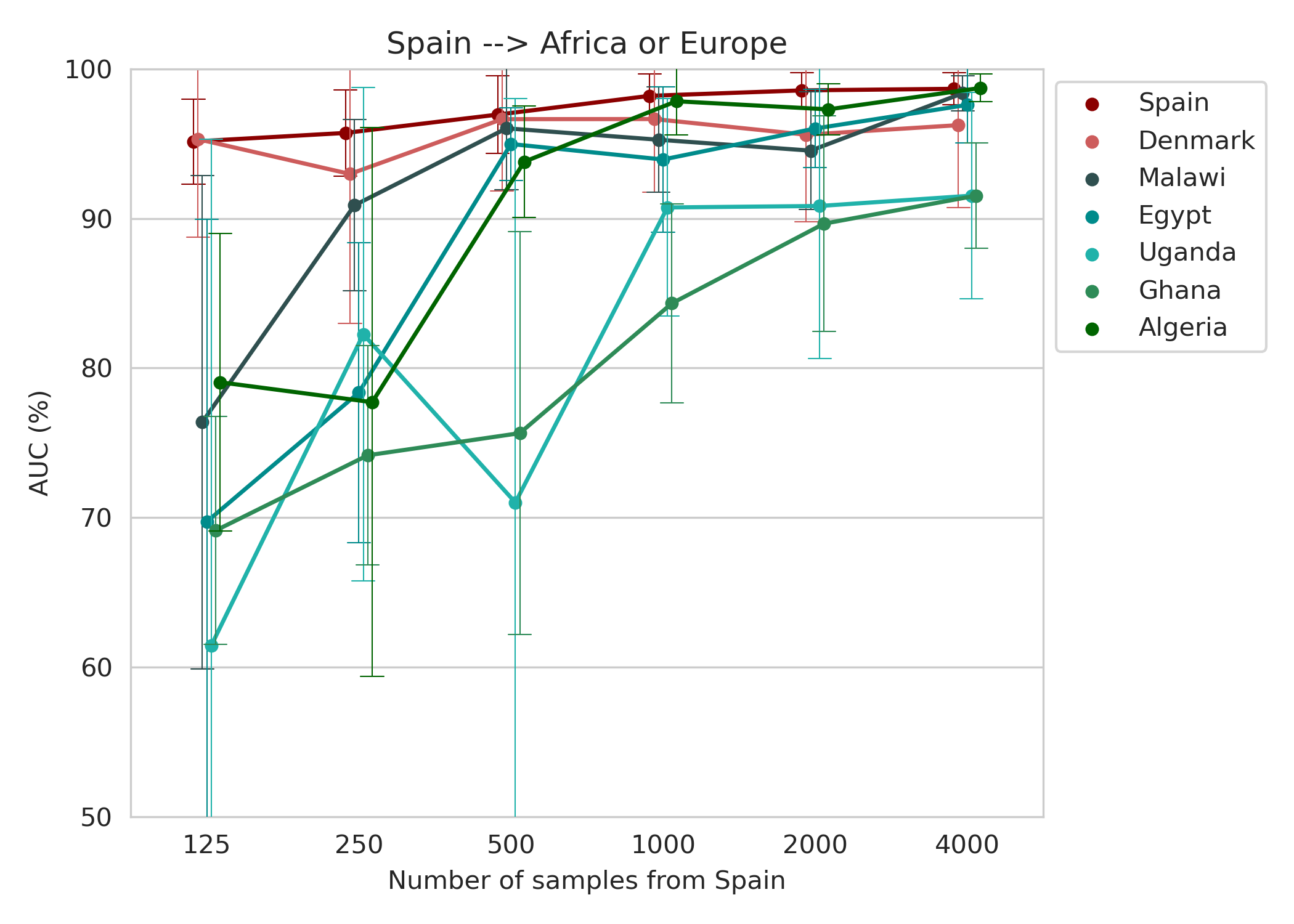}
\caption{Generalisation performance of a model trained with different number of samples from the Spanish dataset evaluated in the same clinical centre, in a new centre in Europe (Denmark) and five datasets from Africa (Malawi, Egypt, Uganda, Ghana, Algeria).}
\label{fig:baseline}
\end{figure}

In the baseline experiment, we evaluated the generalisation of the model trained with the Spanish dataset for different amounts of samples. Figure~\ref{fig:baseline} shows that a good performance is achieved on the Danish dataset (average AUC greater than 95\%) when a sufficiently large amount of samples is used ($n>500$). Although the Spanish model has a good generalisation ability in another European dataset, the African clinical centres with low-resource settings manifest an inconsistent pattern and lower performance, especially when the number of training samples is scarce. In the case where the images are acquired using a similar protocol to that of the European datasets, such as Algeria (Table~\ref{tab:dataset acquisition}), the model shows a competitive performance (average AUC of $97.9\pm1.9$\%) when more than 1000 samples are used for training.

\subsection{Transferability of the fetal plane classifier from high to low-resource settings using a multi-centre approach: combination or transfer learning}

\begin{figure}[h]
\centering
\includegraphics[width=.8\textwidth]{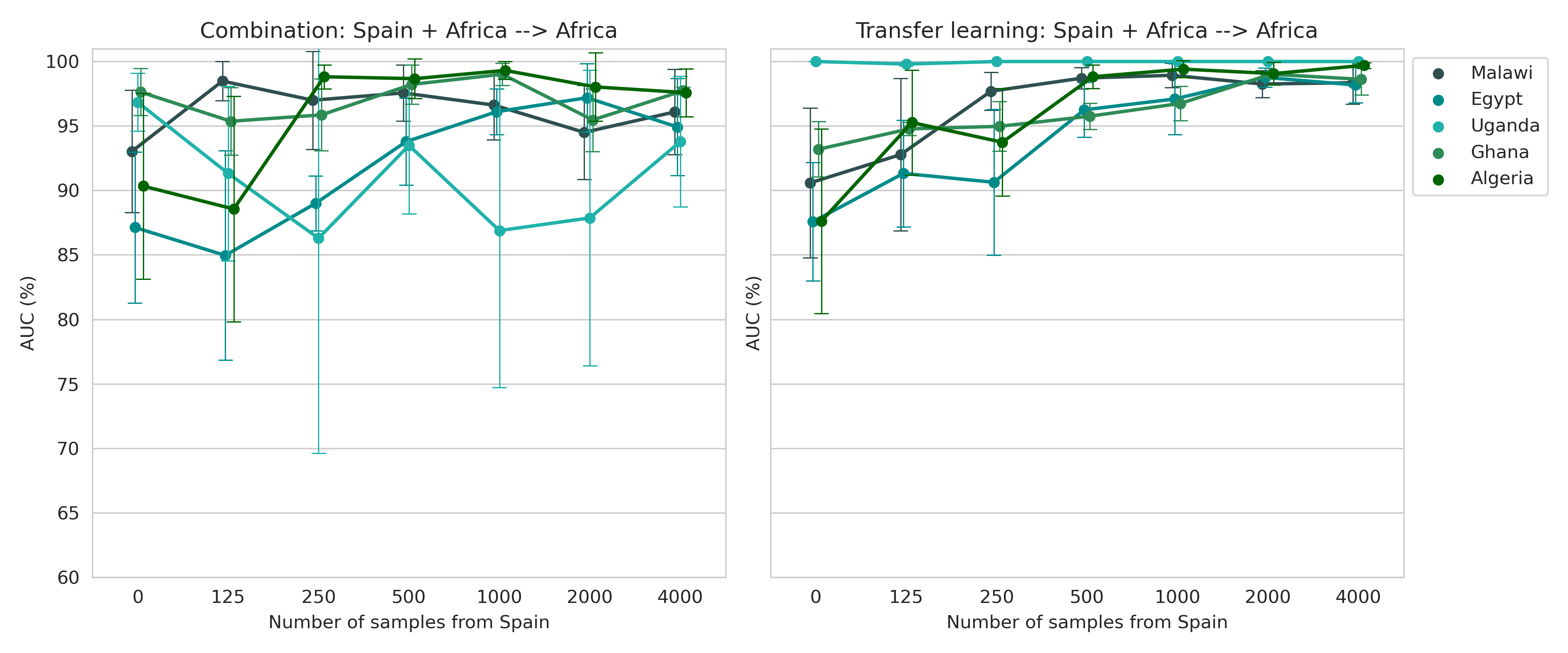}
\caption{The AUC score as a result of combining the dataset acquired in Spain with the US samples acquired in each African centre independently or after fine-tuning the Spanish model, trained with varying sample size, with US images from each African centre independently.}
\label{fig:combination_tl}
\end{figure}

The first approach tested to improve the generalisability of the model from high to low-resource settings consisted of the combination of the Spanish dataset with the small-size datasets from each African centre. The AUC for these models separated by the African dataset used is shown in Figure~\ref{fig:combination_tl} (left plot) for a different number of training samples. The same procedure was followed for the second approach (right plot in Figure~\ref{fig:combination_tl}), which is based on fine-tuning a model, pre-trained with data from Spain, using all patients in Malawi, Egypt, Uganda, Ghana and Algeria. When combining different datasets for training, the number of samples from Spain needed to reach the maximum performance in the African clinical centres is in most cases 500 or 1000. However, this number is not accurate and the good performance of the model is highly dependent on its choice, making it challenging to find a good balance between both datasets. On the other hand, in the transfer learning approach, the performance of the new clinical centres increases progressively with the number of cases from the Spanish dataset. The maximum AUC is obtained when using transfer learning (above 98\% for all target datasets).

In terms of computational cost, Table~\ref{tab:resources} shows that training in the combination approach takes 8 times longer than when using transfer learning. Moreover, although the CPU and model storage is the same, the GPU resources needed are 3 times higher. In summary, the transfer learning approach is the most efficient considering the computational resources required and the final performance.

\renewcommand{\arraystretch}{1.5}
\begin{table}
    \captionof{table}{Summary of the computational resources needed for the combination ($n=500$) and transfer learning approaches ($n=4000$).}
    \scriptsize
    \centering
    \begin{tabular}{|p{3cm}|p{2.5cm}|p{2.5cm}|}
    \hline
     & \textbf{Combination} & \textbf{Transfer learning}\\ 
    \hline
    Inference time (s/epoch) & 8 & 1\\
    \hline
    GPU memory (MB) & 5,683 & 1,803\\
    \hline
    RAM memory (MB) & 6,000 & 6,000\\ 
    \hline
    Model size (MB) & 48.6 & 48.6\\
    \hline
    Batch size & 24 & 24\\
    \hline
    \end{tabular}
\label{tab:resources}
\end{table}

\begin{figure}[h]
\centering
\includegraphics[width=\textwidth]{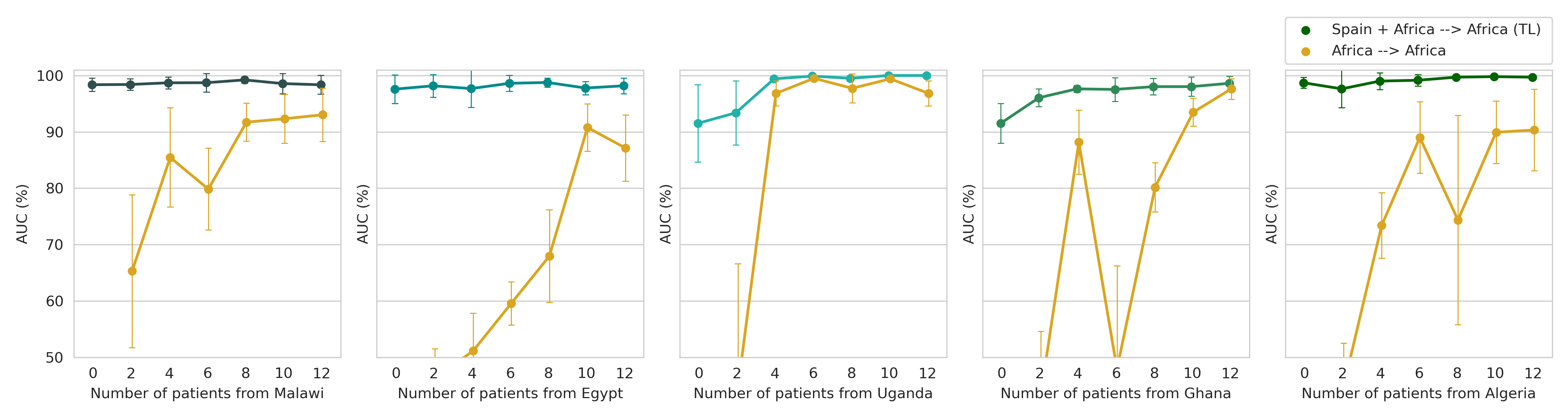}
\caption{The evaluation of the model after being trained in Spain with all available samples and fine-tuned using a different number of patients from each African centre, as compared to the performance of the model directly trained with the African US images.}
\label{fig:tl_min}
\end{figure}

\begin{figure}[h]
\centering
\includegraphics[width=.8\textwidth]{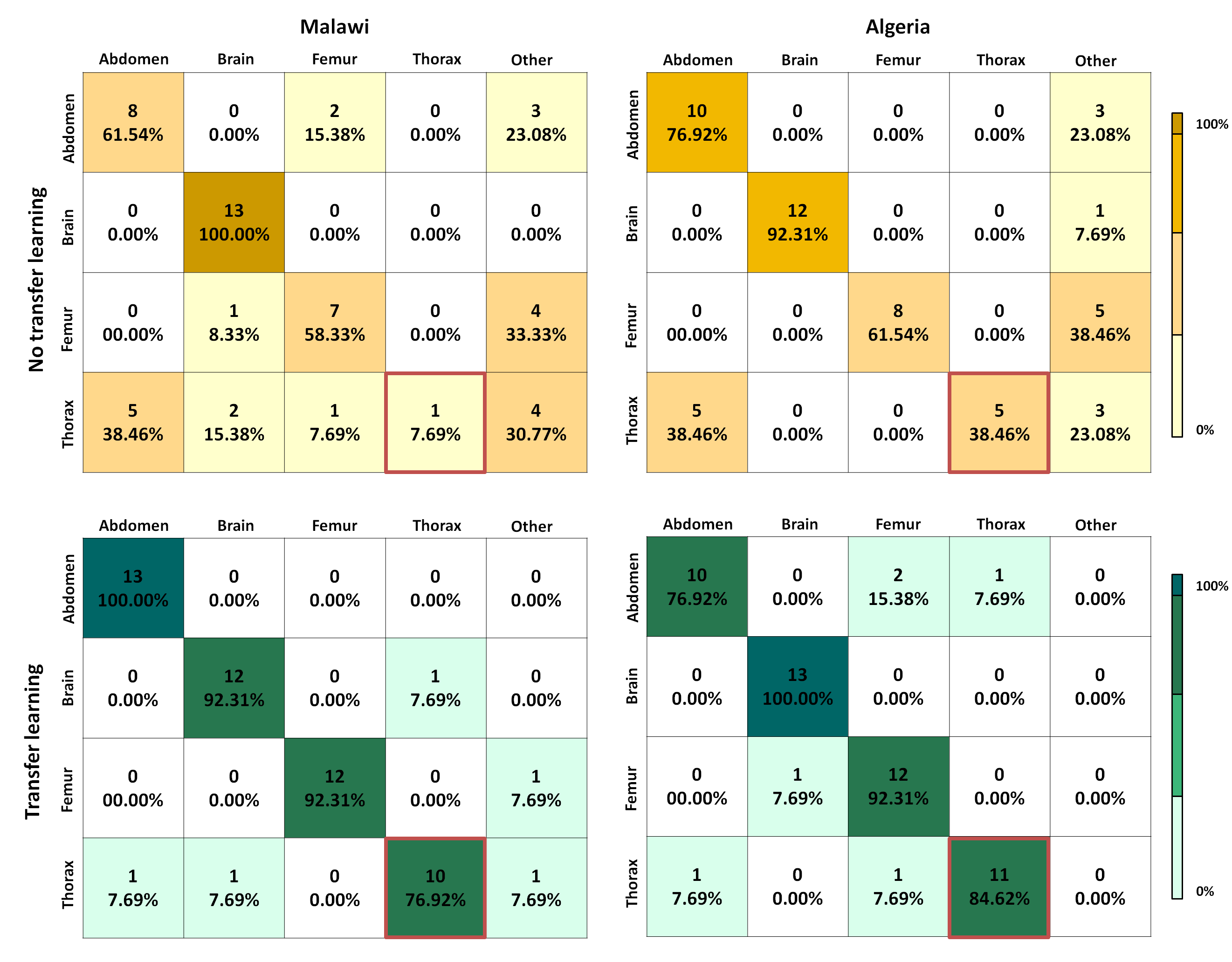}
\caption{Results on common plane classification with and without transfer learning using 12 patients of each African centre in Malawi and Algeria.}
\label{fig:tl_cm}
\end{figure}

Given the potential of the transfer learning approach to result in a generalisable model we now study the optimal number of patients required to get a good performance. Using the model pre-trained with 4,000 images from the Spanish dataset, we obtain in Figure~\ref{fig:tl_min} the AUC for the different African datasets when using a different number of patients from the target domain. We compared this strategy with the case in which a big dataset is not available and so the model has to be trained using only the samples from the African centre. The results indicate that fine-tuning models trained on a large dataset with a small amount of target data is sufficient to reach a high classification performance. The generalisation performance increases progressively with the number of patients used from the target centre and reaches a maximum AUC when using around 4 or 6 patients. On the contrary, when fine-tuning is not used a lower performance is obtained. For example, in the case of Malawi, Egypt and Algeria, the maximum average AUC when training from scratch is 93.3\%, 90.8\% and 90.3\%, respectively, while transfer learning increases the average performance to 99.2\%, 98.8\% and 99.8\%. In Uganda and Ghana, the improvement given by transfer learning is smaller since the maximum baseline performance is already high: 99.5\% and 97.6\% versus 100\% and 98.6\% with fine-tuning, respectively. The higher baseline performance in Ghana and Uganda could be explained by the fact that in these two centres the classes used for testing are only 3 (abdomen, brain and femur). As observed in Figure~\ref{fig:tl_cm}, the thorax plane is more frequently misclassified when fine-tuning is not used. In particular, most of the thorax examples are wrongly classified as brain or `other'. Figure~\ref{fig:tl_cm} also shows that when a model pre-trained in a large dataset is used and fine-tuned, the model can adapt to the new configuration of the new centre.

\begin{figure}[h]
\centering
\includegraphics[width=\textwidth]{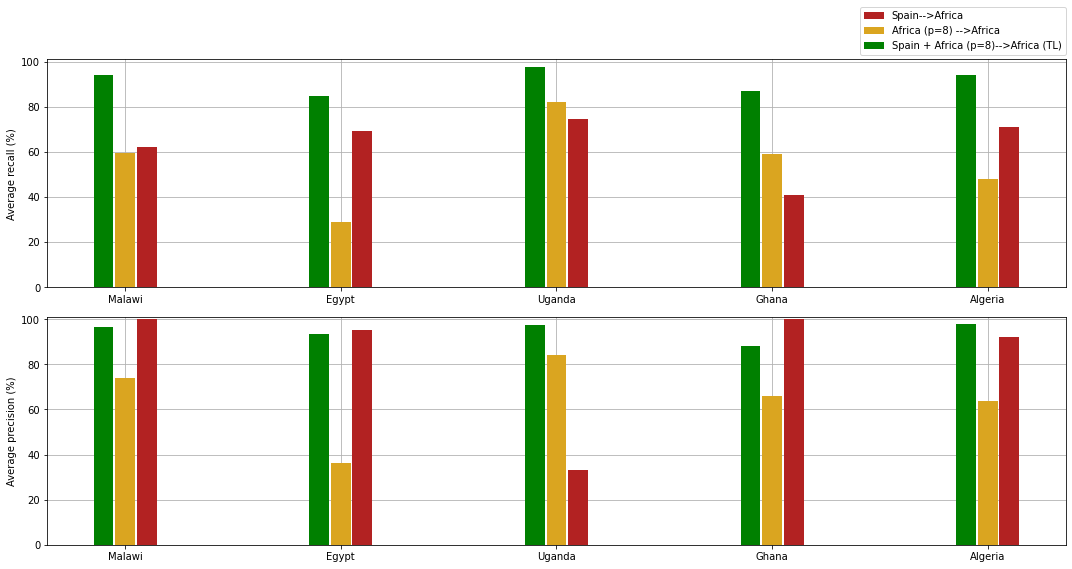}
\caption{A summary of the performance achieved when only 8 patients from Egypt, Uganda, Ghana, Algeria and Malawi are used in terms of recall and accuracy. First, the model is directly trained with the African samples (yellow). Second, the model is trained using only the large Spanish dataset (red). Finally, the pre-trained Spanish model is fine-tuned with the new African samples (green).}
\label{fig:summary}
\end{figure}

Finally, Figure~\ref{fig:summary} quantitatively summarizes the capacity of transfer learning to adjust existing models to new centres with minimal effort and reduced sample size as compared to models trained from scratch with African datasets. Transfer learning achieves an improved average recall, by reducing the number of false negatives, while the average precision is stabilised and reaches a comparable value when compared to models trained with the Spanish dataset.

\section{Discussion}

An important goal for the adoption of deep learning models in the medical domain is to develop inclusive solutions that consider contexts with limited resources. However, to build robust predictive models, neural networks usually require a large number of parameters and large datasets, which are especially difficult to obtain in low-resource settings. For this reason, our work investigated the transferability of deep learning models, developed in centres with greater access to large clinical imaging datasets, to new centres with limited resources, where the generalisation ability is proved to be even more problematic.

Here, we used an existing large-scale data repository from Spain to build a model for US fetal plane classification that can be transferred to a new centre in the African continent. Our results showed that a model trained using a single-centre approach did not reach the desired performance when applied to a new centre in low-resource settings acquired in Malawi, Egypt, Algeria, Uganda or Ghana. In contrast, a multi-centre approach was able to bridge the gap between centres in Africa with less advanced equipment, and centres in Europe with high-quality images and similar acquisition. Specifically, we observed that by directly combining samples from both centres the performance was not as good and stable as with transfer learning, which was capable of effectively adapting the model to the new centre by only fine-tuning the last 4 layers of the model. Through shared representation learning, the neural network has favoured deep phenotypes that are common and consistent across populations/centres, while removing site-specific imaging patterns. With the optimisation of the final weights of the neural network, the objective is to maximise the predictive performance for the African ultrasound datasets, despite their small sample size.

Fine-tuning a model pre-trained with a large imaging dataset in Spain proved to be a potential solution to optimise the performance of a deep learning model in African settings by using small-size datasets and requiring minimal effort. This framework shows promise for generalisability across multi-centre African datasets with challenging and heterogeneous conditions. Moreover, it also encourages the community to conduct further research to develop highly generalisable solutions for the usability of AI in countries with fewer resources and consequently in higher need of clinical support. 

One of the main limitations of the present study is that the datasets obtained in Africa are not complete, since they are less abundant and more difficult to compile. Future works should consider the evaluation of the model on the missing categories.


\section{Conclusions}

This work was motivated by the need for building new AI models generalisable across clinical centres with limited data acquired in challenging and heterogeneous conditions. This framework has shown to be an opportunity for its widespread application to real maternal-fetal clinical settings, especially as a supporting tool in US plane recognition in low-income countries. We believe our work is an important first step in this direction and one that will encourage the development of more generalisable models based on transfer learning in centres with low-resource settings.

\section*{Funding}
This work received funding from the European Union's 2020 research and innovation programme under grant agreement No. 825903 (euCanSHare project), as well as from the Spanish Ministry of Science, Innovation and Universities under grant agreement RTI2018-099898-B-I00. Additionally, the research leading to these results has received funding from Cerebra Foundation for the Brain Injured Child (Carmarthen, Wales, UK).

\section*{Abbreviations}

Artificial Intelligence (AI); Convolutional Neural Network (CNN); General Electric (GE).

\section*{Availability of data and materials}

The Spanish dataset is publicly available at \href{https://zenodo.org/record/3904280}{zenodo.org/record/3904280}. Access to the Danish dataset must be requested to the SONAI project members by contacting Martin G. Tolsgaard at \href{martintolsgaard@gmail.com}{martintolsgaard@gmail.com}. The African datasets will be made available at Zenodo (\href{https://doi.org/10.5281/zenodo.7540448}{10.5281/zenodo.7540448}) upon publication.

\section*{Competing interests}

The authors declare that they have no competing interests.

\section*{Authors' contributions}

Design of the work: CSB, JTB, KL. Image collection: YAA, ME, BOB, PN, WS, MA, LNB, HNK, SGS, MT. Interpretation of results: CSB, JTB, KM, MT, KL. Manuscript draft: CSB, VMC, JTB, KL. Manuscript review: all authors.

\bibliography{bibliography}

\end{document}